\documentclass{appolb}
\usepackage{graphicx}
\usepackage{array}%for math tables
\usepackage{amsmath}%for math tables
\usepackage{amsfonts}
\usepackage{mathbbol}
\usepackage{bm}
\usepackage{tabularx}
\usepackage{tabulary}
\usepackage{dcolumn} % for tables aligned around decimal points

\def\>{\big>}
\def\<{\big<}
\def\|{\big\vert}
\def\£{\big\Vert}
\def\>{\big>}
\def\<{\big<}
\def\){\big)}
\def\({\big(}

\def\etal{\textit{et~al}}

\newcommand{\rf}[1]{(\ref{#1})}

\def\etab{\end{tabular}}  
  
\def\bit{\begin{itemize}}
\def\eit{\end{itemize}}
\def\bml{\begin{multline}}
\def\eml{\end{multline}}
\def\be{\begin{equation}} 
\def\ee{\end{equation}}  
\def\bea{\begin{eqnarray}}  
\def\eea{\end{eqnarray}} 
\def\bmu{\begin{multline}}
\def\emu{\end{multline}}
\def\bal{\begin{array}{l}} 	%for equations which are at least three lines long, 
\def\eal{\end{array}}

\newcommand{\ve}[1]{\mathbf{#1}} 	%roman vector
\def\x{\ve{x}}
\newcommand{\p}{\ve{p}}

\renewcommand{\l}{\lambda}

\def\S{\Sigma}

\def\letterS{S}
\def\letterP{P}
\def\letterD{D}
\def\letterF{F}
\def\greek#1{\def\letter{#1}
\ifx\letter\letterS\Sigma
\else
	\ifx\letter\letterP\Pi
	\else
		\ifx\letter\letterD\Delta
		\else
			\ifx\letter\letterF\Phi
			\else XXX
			\fi
		\fi
	\fi
\fi
}

\def\q{\overline q}
\def\Q{\overline Q}

\def\cc{c\overline c}

\def\sc{s\overline c}
\def\cc{c\overline c}

\def\bb{b\overline b}

\def\bb{b\overline b}

\renewcommand{\u}[1]{\rm{#1}}
\def\cn#1#2#3{^{#1}{\u{#2}}_{#3}}     %normal chemical notation
\def\an{\cn} %atomic notation
\def\sl#1#2{^{#1}{\u{#2}}}    
    
\newcommand{\uS}{\rm{S}}

\def\epem{e^+e^-}

\def\epem{e^+e^-}

\def\nf#1/#2{\dfrac{#1}{#2}}		%normal fraction
\def\sf#1/#2{\sqrt{\nf{#1}/{#2}}}	% sqrt fraction

\def\spato{\mathbf{ O}}

\def\spino{\bm{\chi}}

\def\bsw{\begin{sideways}}
\def\esw{\end{sideways}}

\def\p{_{\phantom{0}}}

\def\qnset#1{\!\bigg[\!\begin{smallmatrix}#1\end{smallmatrix}\!\bigg]}

\def\spread[#1,#2,#3]{\begin{tabularx}{0.2\linewidth}{XXX}$#1$&$#2$ & $#3$\end{tabularx}}

\begin{document}
\title{Meson mass splittings in unquenched quark models (EEF70)\thanks{Presented at \emph{Workshop on Unquenched Hadron Spectroscopy:     Non-Perturbative Models and Methods of QCD vs. Experiment (EEF70).}}%
}
\author{T.~J.~Burns\footnote{\texttt{t.burns@oxon.org}}
\address{Department of Mathematical Sciences, Durham University, DH1 3LE, UK\\
\emph{now at}\\
Department of Physics, Swansea University, SA2 8PP, UK}}
\maketitle
\begin{abstract}
General results are obtained for meson mass splittings and mixings in unquenched (coupled-channel) quark models. Theorems derived previously in perturbation theory are generalised to the full coupled-channel system. A new formula is obtained for the mass splittings of physical states in terms of the  splittings of the valence states. The S-wave hyperfine splitting decreases due to unquenching, but its relation to the vector $e^+e^-$ width is unchanged; this yields a prediction for the missing $\eta_b(3\uS)$. The ordinary (quenched) quark model result that the P-wave hyperfine splitting vanishes also survives unquenching. A ratio of  mass splittings used to discriminate quarkonium potential models is scarcely affected by unquenching.

\end{abstract}
%\PACS{12.39.Jh, 12.39.Pn, 12.40.Yx, 14.40.Pq}
  
\section{Angular momentum coefficients}
Unquenched quark models for meson spectroscopy incorporate $q\q$ pair creation via the transition $Q\Q\to (Q\q)(q\Q)$. Most models have an operator with the same basic structure, and so share the same general solution \cite{Burns:2014zfa,Burns:2013xoa}: this applies to $\an3P0$ models, flux tube models ($\an3P0$ and $\an 3S1$), pseudoscalar-meson emission models, the Cornell model with Lorentz vector confinement and, in the heavy-quark limit, more general microscopic models with Lorentz scalar confinement and one-gluon exchange.

These ``non-flip, triplet'' models are characterised by the assumptions that the initial $Q$ and $\Q$ spins are conserved, and that the created $q\q$ pair is coupled to spin triplet. The operator is a scalar product $\spino\cdot\spato$ of a spin triplet wavefunction $\spino$ (common to all models) and a spatial operator $\spato$ (which differs from model to model). The predictions of such models are consistent with lattice QCD \cite{Burns:2006wz,Burns:2007hk}.
%\begin{figure*}
%\label{topo}
%\includegraphics[width=0.3\linewidth]{topo-eps-converted-to}
%\caption{The $Q\Q\to (Q\q)(q\Q)$ transition responsible for meson strong decay and unquenched quark models.}
%\end{figure*}

In refs \cite{Burns:2014zfa,Burns:2013xoa} a general expression is obtained for the transition matrix element valid for all non-flip, triplet models. The initial $Q\Q$ state is characterised by a radial quantum number $n$, and spin, orbital and total angular momenta $S$, $L$ and $J$. Similarly the $Q\q$ and $q\Q$ mesons have quantum numbers $n_1S_1L_1J_1 $ and $n_2S_2L_2J_2$, are coupled to angular momentum $j$, and are in a partial wave $l$. The matrix element factorises
\be
M_{jl}
\qnset
{n\p S\p L\p J\p\\
n_1S_1L_1J_1\\
n_2S_2L_2J_2}
=
% \sum_{L'l'}
% \xi_{jl}^{L'l'}
% \qnset
% {S\p L\p J\p\\
% S_1L_1J_1\\
% S_2L_2J_2}
% A_{l}^{L'l'}
% \qnset
% {n\p L\p\\
% n_1L_1\\
% n_2L_2}
% =
\bm{\xi}_{jl}
\qnset
{S\p L\p J\p\\
S_1L_1J_1\\
S_2L_2J_2}
\cdot
\bm{A}_{l}
\qnset
{n\p L\p\\
n_1L_1\\
n_2L_2},
\ee
where $\bm \xi$ and $\bm A$ are the matrix elements of $\spino$ and $\spato$ respectively, along with some angular momentum factors. The dependence of the matrix element on the relative momenta of the meson pair is contained in $\bm A$.

The  angular momentum  coefficients $\bm \xi$ are model independent and are discussed in detail in refs. \cite{Burns:2014zfa,Burns:2013xoa}. For present purposes we need only exploit their orthogonality, which leads to the closure relation
\be
 \sum_{\substack{S_1J_1\\S_2J_2\\j}}
 M_{jl}
 \qnset
 {\widehat n\p \widehat S\p \widehat L\p J\p\\
 n_1S_1L_1J_1\\
 n_2S_2L_2J_2}^*
 M_{jl}
 \qnset
 {n\p S\p L\p J\p\\
 n_1S_1L_1J_1\\
 n_2S_2L_2J_2}
=\delta_{\widehat S S}\delta_{\widehat L L}
\bm{A}_{l}^*
\qnset
{\widehat n\p \widehat L\p\\
n_1L_1\\
n_2L_2}
\cdot
\bm{A}_{l}
\qnset
{n\p L\p\\
n_1L_1\\
n_2L_2}
\label{closure}.
\ee

\section{The coupled-channel problem}

The eigenstates $i$ of the coupled-channel problem are admixtures of valence states $Q\Q$ and meson-meson continua $(Q\q)(q\Q)$. In solving for the eigenvalues  $E_i$, the key quantity is the following matrix element
\be
\<\widehat n \widehat S \widehat L J \|\Omega(E_i)\| nSLJ\>
=\sum_{\substack{n_1{S_1}L_1{J_1}\\n_2{S_2}L_2{J_2}\\{j}l}}
\int d p p^2
\frac{M_{jl}
 \qnset
 {\widehat n\p \widehat S\p \widehat L\p J\p\\
 n_1S_1L_1J_1\\
 n_2S_2L_2J_2}^*
 M_{jl}
 \qnset
 {n\p S\p L\p J\p\\
 n_1S_1L_1J_1\\
 n_2S_2L_2J_2}}{E_{12}(p)-E_i},
\ee
where $p$ and $E_{12}$ are the momenta and energy of the contiuum mesons. 

If there are no spin splittings among the continua, the closure relation~\rf{closure} can be exploited. (This is otherwise not possible since $E_{12}$ depends on $S_1$, $J_1$, $S_2$ and $J_2$ through the continuum meson masses.) This gives
\bea
\<\widehat n \widehat S \widehat L J \|\Omega(E_i)\| nSLJ\>&=&\delta_{\widehat S S}\delta_{\widehat L L}\<\widehat n L \|\Omega(E_i)\| nL\>,\textrm{ with}
\\
\<\widehat n L \|\Omega(E_i)\| nL\>
&=&
\sum_{\substack{n_1L_1\\n_2L_2\\l}}
\int dp p^2
\frac{\bm{A}_{l}^*
\qnset
{\widehat n\p  L\p\\
n_1L_1\\
n_2L_2}
\cdot
\bm{A}_{l}
\qnset
{n\p L\p\\
n_1L_1\\
n_2L_2}
}{E_{12}(p)-E_i}.
\eea
In this approximation there is no mixing due to unquenching among states with different $S$ or $L$. This is a generalisation of a theorem obtained in perturbation theory \cite{Barnes:2007xu} to the full coupled-channel problem.

If we further assume (and this assumption will be relaxed shortly) that there are no spin splittings among the valence masses, then since the mixing matrix is independent of $S$ and $J$, the physical masses are also independent of $S$ and $J$ (another generalisation of ref. \cite{Barnes:2007xu}), as is the configuration mixing of different radial states (a new result).

We return now to the more general case with splittings among the valence and (consequently) physical masses, but not among the continua. (Small continuum splittings can be dealt with, and do not modify the results below.) Ignoring mixing among different radial states, the physical mass $E_{nSLJ}$ of a state below threshold is related to its valence mass $M_{nSLJ}$,
 \be
 E_{nSLJ}=  M_{nSLJ}-\<\Omega (E_{nSLJ}) \>_{nSLJ}
  \ee
and the squared amplitude that the state is in the valence configuration is
 \be
  Z_{nSLJ}=\frac{1}{1+\<\omega(E_{nSLJ})\>_{nSLJ}}\textrm{,\quad with}\quad
 \omega(E_{nSLJ})=\frac{\partial \Omega (E_{nSLJ})}{\partial E_{nSLJ}}.
  \ee
With the following parametrisation of masses
\be
 M_{nSLJ}=M_{nL}+\delta M_{nSLJ},\quad\quad
 E_{nSLJ}=E_{nL}+\delta E_{nSLJ},
\ee
the Taylor expansion of the mass shift about $E_{nL}$ can be written
%\be
%\Omega(E_{nSLJ})\approx \Omega(E_{nL})+\delta E_{nSLJ}\omega(E_{nL})
%\ee
%and assuming that there are no splittings among the continua, %then the matrix elements are independent of $S$ and $J$ (the closure relation),
\be
\<\Omega (E_{nSLJ}) \>_{nSLJ}\approx \<\Omega (E_{nL}) \>_{nL}+\delta E_{nSLJ}\<\omega (E_{nL}) \>_{nL}
\ee
%Separating out the spin-averaged mass shift
%\be
%E_{nL}=M_{nL}-\<\Omega (E_{nL}) \>_{nL}
%\ee
which leads to a relation between the physical and valence spin splittings
\be
{\delta E_{nSLJ}=Z_{nL}\delta M_{nSLJ}}\textrm{,\quad with}\quad
Z_{nL}=\frac{1}{1+\<\omega(E_{nL})\>_{nL}}.
\label{formula}
\ee
This is the main result of this work. Unquenching reduces spin splittings, such	 that the physical splittings are suppressed with respect to the valence splittings by the (spin-averaged) valence component $Z_{nL}$.

% \begin{frame}{Mesons: degenerate continua, split valence states}
% The key result is the relation
% \be
% \delta E_{nSLJ}=Z_{nL}\delta M_{nSLJ}
% \ee
% \bit
% \item unquenching reduces spin splittings, and
% \item the physical splittings are renormalised proportional to the (spin-averaged) valence component.
% \eit
% 
% \bbl{}
% The ``squeezing'' of spin splittings has been observed in $\psi'-\eta_c'$, and improves the fit to data\footnote{\scriptsize e.g. \bibentry{Martin:1982nw}, \bibentry{Eichten:2004uh}}.
% \ebl
% 
% \bbl{}
% But what if we incorporate spin splittings among the continua?
% \ebl
% \end{frame}

The validity of the formula can be checked by comparing its predictions to existing model calculations in the literature. Table \ref{s-wave} gives a typical example, for the hyperfine splitting of S-wave charmonia and bottomonia. A forthcoming paper will discuss the result in more detail, testing its predictions more widely against the literature (including for orbitally excited states). The rest of this Proceedings is instead devoted to some applications.

\begin{table}
	\qquad \qquad\begin{tabular}{l D{.}{.}{-1}	D{.}{.}{-1} 	D{.}{.}{-1}	>{$}r<{$}}
	&\multicolumn{1}{c}{$\<\Omega\>_{n\uS}$}	
		&\multicolumn{1}{c}{$\delta M$}
			&\multicolumn{1}{c}{$\delta E$}
				&\multicolumn{1}{c}{$\delta E^{pred.}$}	\\
\hline
$\cc$ \cite{Kalashnikova:2005ui}&&&&\\
\hline
1S 	&174	&129	&117	&{116.4}	\\
2S 	&212	&64	&48	&{48.4}		\\
\hline
 $\bb$ \cite{Liu:2011yp} &&&&\\
 \hline
1S 	&57.41	&71.39	&68.50	&^*{68.44}	\\
2S 	&67.58	&23.12	&21.30	&{21.36}	\\
3S 	&67.74	&15.73	&14.00	&{14.06}	\\
 \hline
\end{tabular}
\caption{Some model calculations (from refs \cite{Kalashnikova:2005ui,Liu:2011yp}) for the spin-averaged mass shifts $\<\Omega\>_{nS}$, the bare and physical hyperfine splittings $\delta M$ and $\delta E$, and the splittings $\delta E^{pred.}$ predicted by equation \rf{formula}. All quantities are in MeV. The entry~($^*$) uses the author's own calculation for $Z_{n\uS}$, which disagrees with ref. \cite{Liu:2011yp}.}
\label{s-wave}
\end{table}

\section{Some applications}

\subsection{S-wave hyperfine splitting and $\epem$ widths}
In the quenched quark model, meson hyperfine splittings and $\epem$ widths are both proportional to the square of the $Q\Q$ wavefunction at the origin, which leads to the model-independent relation:
 \be
\delta M_{2\uS}/\delta M_{1\uS}=\Gamma_{\epem\to 2\an3S1}/\Gamma_{\epem\to 1\an3S1}
 \ee
 The relation is satisfied by experimental data for charmonia and bottomonia,
% \begin{align}
% \frac{E_{\psi'}-E_{\eta_c'}}{E_{J/\psi}-E_{\eta_c}}&=0.407\pm 0.015,
% & \frac{\Gamma_{\epem\to \psi'}}{\Gamma_{\epem\to J/\psi}}&=0.423\pm 0.018\\
% \frac{E_{\Upsilon'}-E_{\eta_b'}}{E_\Upsilon-E_{\eta_b}}&=0.42\pm0.09, 
%& \frac{\Gamma_{\epem\to \Upsilon'}}{\Gamma_{\epem\to \Upsilon}}&=0.457\pm 0.014
% \end{align}
so it is important to establish that it survives the effects of unquenching~\cite{Burns:2012pc}. Unquenching suppresses the physical mass splittings by a factor $Z_{n\uS}$, but at the same time, suppresses the $\epem$ widths by $Z_{n\an 3S1}$ (assuming that they are dominated by the $Q\Q$ component). To a very good approximation $Z_{n\an3S1}\approx Z_{n\uS}$, so the relation survives with physical masses. The corresponding relation between the 1S and 3S levels yields a mass prediction  $10334.6\pm 2.2$ MeV for  the $\eta_b(3\uS)$.
% \begin{frame}{Hyperfine splitting ($L=0$): prediction for $\eta_b(3S)$}
% We can use the relation to predict the 3S hyperfine splitting of $\bb$
% \be
% \delta M_{3S}=20.6 \pm 1.7~MeV
% \ee
% and the $\eta_b(3S)$ mass
% \be
% M_{\eta_b(3S)}=10334.6\pm 2.2~MeV
% \ee
% \end{frame}
\subsection{P-wave hyperfine splitting}
The quenched quark model result for the  P-wave hyperfine splitting 
\be
\tfrac{1}{9}\left(M_{\an 3P0}+3M_{\an 3P1}+5M_{\an 3P2}\right)-M_{\an 1P1}= 0
\ee
is satisfied by charmonia and bottomonia. Since each of the states in the multiplet is subject to large, and different, mass shifts, \emph{a priori} the result could be spoiled by the effects of unquenching. Remarkably, across most models these shifts conspire to make very little contribution to the hyperfine splittings; for example, the shifts (MeV) of 1P bottomonia from ref.\cite{Liu:2011yp} give:
\be
\tfrac{1}{9}\left(80.777+3\times 84.823 + 5\times 87.388\right)-85.785=0.013
\ee
The vanishing hyperfine splitting is protected by a mechanism observed and explained in ref. \cite{Burns:2011jv,Burns:2011fu}, but can also be seen as a simple consequence of equation \rf{formula}: if the bare states have zero hyperfine splitting, so too do the physical states. Corrections to \rf{formula} due to different continuum masses turn out to affect only the spin-orbit splittings at leading order. The vanishing D-wave splitting (also protected by this mechanism) leads to a prediction for the missing $\an 1D2$ bottomonium \cite{Burns:2010qq}.

\subsection{The $Q\Q$ potential}

The ratio of mass splittings $ R=({M_{\an3P2}-M_{\an3P1}})/({M_{\an3P1}-M_{\an3P0}})$ has been used by many authors to discriminate among models for the $Q\Q$ potential, which raises the question of whether such conclusions should be modified due to unquenching. According to equation \rf{formula}, the ratio $R$ is invariant under unquenching, and so conclusions based on quenched quark models survive. In practice there are corrections to equation \rf{formula} which lead to a decrease of $R$ due to unquenching, but the effect is not substantial.

%\begin{table}
%\begin{tabular}{llllll}
%\hline
%$\cc$	& $R^{(bare)}$	&$R^{(phys.)}$	\phantom{xxxxxxxxx}
%			&$\bb$	&$R^{(bare)}$	&$R^{(phys.)}$\\
%\hline
%1P (LMC)	&0.65	&0.57	&LD (1P)	&0.736	&0.723	\\
%1P (K)	&0.51	&0.47	&LD (2P)	&0.741	&0.722	\\
%1P (FS)	&0.59	&0.51	&LD (3P)	&0.753	&0.705	\\
%1P (OT)	&0.44	&0.42	&OT (1P)	&0.913	&0.909	\\
%1P (YLCD)	&0.59	&0.47	&OT (2P)	&0.894	&0.833	\\
%\hline
%\end{tabular}
%\end{table}
\subsection{Leptonic width inequalities}
Expanding $\omega(E_{nSLJ})$ in a similar way to $\Omega(E_{nSLJ})$ leads to inequalities among the $Z$-factors, for example
$ Z_{^1S_0}>Z_{^3S_1}$ and $Z_{^3P_0}>Z_{^3P_1}>Z_{^3P_2}$.
These modify relations among leptonic widths which arise due to eliminating common factors of the wavefunction at the origin, such as 
\be
\frac{\Gamma_{^1S_0\to\gamma\gamma}}{\Gamma_{^3S_1\to\epem}}>\frac{4}{3}\left(1+1.96\frac{\alpha_s}{\pi}\right),\qquad
\frac{\Gamma_{^3P_2\to\gamma\gamma}}{\Gamma_{^3P_0\to\gamma\gamma}}<\frac{4}{15}\left(1-5.51\frac{\alpha_s}{\pi}\right).
\ee
It will be difficult to identify such effects in practice, as leptonic width relations are subject to large theoretical and experimental uncertainties; nevertheless it is instructive to know the direction in which unquenching modifies such relations.

\subsection{Splittings in lattice QCD}
In unquenched lattice QCD, spin splittings vary with dynamical quark masses. On the basis of equation \rf{formula}, one might expect that decreasing the quark mass leads necessarily to a decrease in spin splittings. (Due to a decrease in binding energy and an enhancement in the coupling matrix element, the valence component would decrease.) The situation is not so simple, though, since the quark mass influences $\alpha_s$, and moreover there is considerable evidence that unquenched lattice QCD without explicit $(Q\q)(q\Q)$ operators is not sensitive to the coupling $Q\Q\to (Q\q)(q\Q)$ \cite{Thomas:2011rh}.

Nevertheless future lattice calculations with (complete multiplets of) $(Q\q)(q\Q)$ operators could in principle offer a direct test of equation \rf{formula}, if spin splittings and $Z$-factors are measured at various quark masses. There is already some work in this direction; Bali \etal.\cite{Bali:2011rd} have measured splittings and $Z$-factors for several charmonia, but at one quark mass and with one  $(Q\q)(q\Q)$ operator per channel.

% \begin{frame}{Splittings in lattice QCD}
% 
% In principle a lattice QCD calculation with operators
% \bit
% \item $Q\Q$ and 
% \item $(Q\q)(q\Q)=P\o P, P \o V, V \o P, V\o V$
% \eit
% could check the relation
% \be
% \delta E_{nSLJ}=Z_{nL}\delta M_{nSLJ}
% \ee
% by measuring splittings and Z-factors at various quark masses. This would be a test of the $\spino\cdot\spato$ coupling.
% \bbl{}
% Bali \etal.\footnote{\scriptsize\bibentry{Bali:2011rd}} have a single data point for each of several $\cc$, with a limited set of continua.
% \ebl
% \end{frame}

\section{Conclusion}

General results have been obtained for unquenched quak models based on the non-flip, triplet operator. Previous results from perturbation theory, valid in the absence of spin splittings, have been generalised to the full coupled-channel problem, and extended. The more realistic scenario, incorporating spin splittings among the valence and physical masses, involves a simple mass formula, some of whose implications have been discussed here. The formula ensures that several empirically successful results of quenched quark models survive the effects of unquenching. The formula should be testable in future lattice QCD calculations. Although it was not discussed here, the formula is also useful for practical calculations of mass splittings in unquenched quark models.
\bibliography{//tawe_dfs/users_staff/SFS1/T.Burns/Documents/physics/bibinputs/tjb}
\end{document}